# Decentralized Intelligence in GameFi: Embodied AI Agents and the Convergence of DeFi and Virtual Ecosystems

**Fernando Jia**[1,2,*], **Jade Zheng**[2] and **Florence Li**[3]

[1]UC Berkeley RDI,
[2]Starward Games,
[3]Stanford University
*Corresponse: fernando.jia@berkeley.edu

## Abstract

In the rapidly evolving landscape of GameFi, a fusion of gaming and decentralized finance (DeFi), there exists a critical need to enhance player engagement and economic interaction within gaming ecosystems. Our GameFi Ecosystem aims to fundamentally transform this landscape by integrating advanced embodied AI agents into GameFi platforms. These AI agents, developed using cutting-edge large language models (LLMs), such as GPT-4 and Claude AI, are capable of proactive, adaptive, and contextually rich interactions with players. By going beyond traditional scripted responses, these agents become integral participants in the game's narrative and economic systems, directly influencing player strategies and in-game economies.

We address the limitations of current GameFi platforms, which often lack immersive AI interactions and mechanisms for community engagement or creator monetization. Through the deep integration of AI agents with blockchain technology, we establish a consensus-driven, decentralized GameFi ecosystem. This ecosystem empowers creators to monetize their contributions and fosters democratic collaboration among players and creators. Furthermore, by embedding DeFi mechanisms into the gaming experience, we enhance economic participation and provide new opportunities for financial interactions within the game.

Our approach enhances player immersion and retention and advances the GameFi ecosystem by bridging traditional gaming with Web3 technologies. By integrating sophisticated AI and DeFi elements, we contribute to the development of more engaging, economically robust, and community-centric gaming environments. This project represents a significant advancement in the state-of-the-art in GameFi, offering insights and methodologies that can be applied throughout the gaming industry.

## 1 Introduction

The advent of blockchain technology has catalyzed the emergence of decentralized finance ( DeFi) and non-fungible tokens (NFTs), reshaping traditional financial systems and digital asset ownership (Schär, 2021; Chen & Bellavitis, 2020). Within this transformative landscape, GameFi, a fusion of gaming and DeFi, has gained significant traction by introducing financial mechanisms into gaming environments, allowing players to earn real economic value through gameplay. GameFi platforms leverage blockchain transparency and security to empower players with true ownership of in-game assets, fostering play-to-earn models that have redefined user engagement and monetization strategies (Zyda, 2005).

Despite promising prospects, maintaining long-term player participation and enhancing the depth of economic interactions within GameFi platforms remain critical challenges. Traditional GameFi applications often lack sophisticated artificial intelligence (AI) capabilities, resulting in static and predictable gameplay that fails to adapt to individual player behaviors or provide personalized experiences. This limitation not only diminishes player retention but also restricts the potential for dynamic in-game economies that mirror the complexities of real-world financial systems (Sifa, Bauckhage, & Drachen, 2014).

Simultaneously, advancements in AI, particularly in large language models (LLMs) like GPT-3 and GPT-4, have demonstrated remarkable capabilities in natural language understanding and generation (Brown et al., 2020; Achiam et al., 2023). Embodied AI agents—agents capable of interacting within virtual or physical environments—offer new avenues for creating immersive, responsive gaming experiences (Mott & Lester, 2006; Savva et al., 2019). These agents can learn from and adapt to user interactions, providing a level of dynamism and personalization previously unattainable in-game environments (Bakkes, Spronck, & Van den Herik, 2009).

However, integrating advanced AI agents into decentralized gaming platforms presents significant technical and infrastructure challenges. Issues such as scalability, performance bottlenecks, and maintaining decentralization principles complicate the implementation of AI within blockchain-based games (Xie et al., 2019; Zheng, Xie, Dai, Chen, & Wang, 2018). Moreover, current GameFi platforms often lack mechanisms to effectively support community participation and creator monetization, limiting collaborative content creation and innovation within the ecosystem (Kow & Nardi, 2010).

## 1.1 Problem Definition

**Lack of Immersive AI Interactions in GameFi Platforms.** Existing GameFi platforms predominantly feature deterministic game mechanics and limited AI functionality, leading to repetitive gameplay experiences (Sifa et al., 2014). The absence of advanced AI agents restricts the game's ability to adapt dynamically to player actions, hindering personalization and long-term engagement. This limitation is evident in popular GameFi titles where economic incentives are provided, but the gameplay lacks depth and adaptability.

**Insufficient Community Engagement and Creator Monetization Mechanisms.** While blockchain technology facilitates decentralized ownership, many GameFi platforms do not fully leverage this potential to empower creators and communities (Xie et al., 2019). There is a significant gap in the provision of tools and systems that allow creators to contribute content and monetize their efforts effectively (Kow & Nardi, 2010). This deficiency limits the diversity of content and stifles community-driven innovation, which is essential for a vibrant gaming ecosystem.

**Technical Challenges in Integrating Advanced AI into Decentralized Environments.** Integrating AI agents into decentralized platforms is fraught with challenges related to scalability, security, and maintaining decentralization (Zheng et al., 2018; Chen & Bellavitis, 2020). AI models, particularly LLMs, require substantial computational resources, which conflict with the resource constraints inherent in blockchain networks. Ensuring that AI functionalities do not compromise the decentralized ethos of blockchain platforms is a complex issue that has not yet been fully addressed (Nguyen & Kim, 2018).

**Underutilization of DeFi Mechanisms to Enhance Economic Participation.** Although DeFi has introduced innovative financial instruments, their integration within gaming platforms is often superficial (Schär, 2021; Chen & Bellavitis, 2020). GameFi applications sometimes offer token-based rewards without incorporating deeper DeFi functionality such as stakes, yield farming, or decentralized governance, limiting economic participation and opportunities for players and creators.

## 1.2 Importance of Addressing These Challenges

**Enhancing Player Engagement and Retention.** Addressing the lack of immersive AI interactions is crucial to increase player engagement on GameFi platforms. By incorporating advanced AI agents, games can offer personalized experiences that adapt to individual player behavior, improving satisfaction and retention rates (Bakkes et al., 2009). Studies have shown that adaptive AI can significantly improve the user experience by providing appropriate challenges and interactions.

**Fostering Community Development and Creator Empowerment.** Implementing mechanisms for community engagement and creator monetization can lead to a more vibrant and innovative GameFi ecosystem. Empowering creators with tools and economic incentives encourages the development of diverse content, fostering an active community that contributes to the growth of the platform (Kow & Nardi, 2010).

**Advancing the Integration of AI and DeFi Technologies.** By overcoming technical challenges associated with integrating AI into decentralized environments, it is possible to create more sophisticated and scalable GameFi platforms. This advancement not only enhances the gaming experience but also pushes the boundaries of what is possible with blockchain technology, contributing to the broader field of decentralized applications (Savva et al., 2019).

**Enhancing Economic Participation Through Advanced DeFi Mechanisms.** Integrating deeper DeFi functionalities into GameFi can unlock new economic opportunities for players and creators, leading to more robust in-game economies. Features like staking, governance tokens, and decentralized exchanges can increase user engagement and provide additional incentives for participation (Schär, 2021).

## 1.3 Key Unique Contributions

Our project aims to address these multifaceted challenges through the following unique contributions:

**Development of Advanced Embodied AI Agents.** We use state-of-the-art LLMs, such as GPT-4, to develop embodied AI agents capable of sophisticated interactions (Achiam et al., 2023; Brown et al., 2020). These agents can understand and generate human-like responses, adapt to player behaviors, and provide personalized experiences. By integrating reinforcement learning techniques, agents continually improve from interactions within the game environment (Sutton, 2018; Silver et al., 2016).

**Seamless Integration into Decentralized GameFi Ecosystems.** Our approach involves embedding AI agents into blockchain-based gaming platforms using smart contracts and decentralized protocols (Antonopoulos, 2014; Xie et al., 2019). We ensure that AI functionalities are compatible with the decentralized infrastructure, maintaining security and transparency. By utilizing off-chain computation and Layer 2 scaling solutions, we address scalability issues associated with running AI models on blockchain networks (Nguyen & Kim, 2018).

**Empowering Creators with Monetization Opportunities.** We provide creators with tools to develop content using embodied AI agents and monetize their efforts within the GameFi ecosystem (Kow & Nardi, 2010). Our platform supports the creation and trading of NFTs linked to AI-generated content, enabling creators to earn from their contributions. We implement decentralized governance models that allow creators to have a say in platform development and policies (Wright & De Filippi, 2015).

**Enhancing Economic Participation Through Integrated DeFi Mechanisms.** Our platform integrates advanced DeFi tools, including staking, liquidity pools, yield farming, and governance tokens (Schär, 2021). AI agents assist players in navigating these financial instruments, making DeFi more accessible and engaging within the gaming context. This integration fosters a robust in-game economy where players can actively participate and benefit economically.

**Bridging Traditional Gaming with Web3 Technologies Using Embodied AI Agents.** We aim to create a bridge between traditional gaming and Web3 technologies by integrating embodied AI agents into existing game environments. Our plugins allow for the addition of AI interactions and Web3 capabilities in a decentralized manner, facilitating the adoption of blockchain technologies in mainstream gaming. This approach democratizes game development and promotes collective growth within the gaming community (Batubara, Ubacht, & Janssen, 2018; Antonopoulos, 2014).

## 2 Related Works

### 2.1 Decentralized Oracle Networks and Data Provenance in Blockchain Applications

Decentralized Oracle Networks (DONs) play a critical role in enabling smart contracts to interact with off-chain data in a secure and reliable manner. DECO (short for decentralized oracle) is a notable protocol that leverages TLS to allow provers to prove the provenance of web data without requiring server-side modifications or trusted hardware (Zhang, Maram, Malvai, Goldfeder, & Juels, 2020). DECO addresses the challenge of securely exporting authenticated data from websites by introducing a three-phase protocol that includes a three-party handshake, query execution, and proof generation. This protocol ensures that the prover cannot forge session data, and it allows for efficient zero-knowledge proofs of data provenance and fine-grained statements about session content.

Chainlink 2.0 further advances DON capabilities by providing the infrastructure for securely integrating AI and off-chain computation with smart contracts (Breidenbach et al., 2021). By combining on-chain and off-chain resources, Chainlink's DONs facilitate hybrid smart contracts that can support complex and scalable applications, such as those required in GameFi ecosystems.

### 2.2 GameFi: The Intersection of Gaming, DeFi, and Oracles

GameFi represents the fusion of gaming and decentralized finance (DeFi), leveraging blockchain technology to create gaming environments where players can earn real economic value (Schär, 2021). By integrating blockchain's capabilities for secure, transparent ownership and transfer of digital assets, GameFi platforms empower players with true ownership of in-game items, often represented as non-fungible tokens (NFTs), and enable play-to-earn models (Zyda, 2005).

However, sustaining long-term player engagement and fostering robust in-game economies remain significant challenges (Sifa et al., 2014). Many existing GameFi platforms offer limited interactive experiences and lack sophisticated artificial intelligence (AI) capabilities, resulting in static gameplay that does not adapt to the behaviors of individual players (Bakkes et al., 2009).

### 2.3 Decentralized Finance (DeFi), Blockchain, and Oracles in Gaming

Decentralized finance introduces innovative financial instruments and systems built on blockchain technology, aiming to create open, transparent, and accessible financial services (Schär, 2021; Chen & Bellavitis, 2020). In the gaming context, blockchain enables secure and transparent transactions, while DeFi mechanisms introduce economic incentives and financial functionalities within games (Schär, 2021). A critical component of integrating DeFi into gaming is the use of oracles, which provide external data to smart contracts.

DECO's approach to decentralized oracles allows users to prove that data accessed via TLS came from a particular website and optionally prove statements about such data in zero-knowledge, without requiring server-side modifications or trusted hardware (Zhang et al., 2020). This capability is essential for GameFi platforms that need to securely verify data from external sources.

Chainlink has been pivotal in the development of decentralized oracle networks (DONs) that enhance and extend the capabilities of smart contracts by providing connectivity to real-world data, off-chain computation, and secure external communication (Breidenbach et al., 2021). The introduction of Chainlink 2.0 outlines a vision in which DONs can support hybrid smart contracts by combining on-chain and off-chain resources, enabling more complex and scalable gaming applications.

Automated market makers (AMM) and decentralized exchanges (DEXs) facilitate decentralized trading without traditional intermediaries (Angeris, Kao, Chiang, Noyes, & Chitra, 2021). Integrating such DeFi mechanisms into games can revolutionize in-game economies by providing seamless and trustless exchanges of virtual assets (Egorov, 2019). However, challenges such as trade execution costs, bid-ask spreads, and price slippage must be addressed within the gaming context (Bessembinder & Venkataraman, 2010; Qin, Zhou, Livshits, & Gervais, 2021; Qin, Zhou, Gamito, Jovanovic, & Gervais, 2021).

Current GameFi platforms often integrate basic DeFi features, such as token staking and NFT trading, but typically lack deeper DeFi functionalities such as yield farming, liquidity provision, and decentralized governance (Schär, 2021). Furthermore, these platforms frequently do not incorporate advanced AI-driven gameplay, limiting player engagement and the richness of in-game economies (Sifa et al., 2014).

### 2.4 Artificial Intelligence and Embodied AI Agents in Gaming

Artificial intelligence has played a significant role in improving gaming experiences by providing dynamic and adaptive gameplay (Bakkes et al., 2009). Embodied AI agents, capable of interacting within virtual environments, offer opportunities to create more immersive and personalized experiences (Mott & Lester, 2006; Savva et al., 2019).

Bakkes et al. (2009) discussed methods for rapid and reliable adaptation of video game AI to improve player satisfaction. Their work emphasizes the importance of AI that can adjust to player behaviors, providing appropriate challenges and enhancing engagement. Similarly, Mott et al. (2006) explored narrative-centered tutorial planning using AI agents to create engaging and educational game experiences.

Advancements in AI, particularly in large language models (LLMs) such as GPT-3 and GPT-4, have demonstrated the potential of AI to generate human-like text and interactions (Brown et al., 2020; Achiam et al., 2023). These models can be used to develop AI agents with sophisticated natural language processing capabilities, which enhances interaction within games.

## 2.5 Integration of AI Agents with Blockchain, GameFi, and Oracles

The integration of AI agents into blockchain-based gaming environments is an emerging field. Previous research has explored the use of AI in gaming and the development of DeFi applications on blockchain separately, but the convergence of these domains presents new challenges and opportunities.

Chainlink's vision for DONs includes providing the necessary infrastructure for securely integrating AI and off-chain computation with smart contracts (Breidenbach et al., 2021). By enabling decentralized off-chain computation and off-chain data storage, Chainlink's DONs can facilitate the deployment of AI agents within GameFi ecosystems, enhancing scalability, confidentiality, and performance.

The DECO approach to decentralized oracles can also be instrumental in integrating AI with blockchain by securely proving the provenance and properties of the online data without requiring server-side cooperation (Zhang et al., 2020). This capability is crucial when AI agents need to interact with external data sources.

Automated trading agents and AMMs in DeFi have been analyzed for their market microstructure and impact on liquidity and trading costs (Biais, Glosten, & Spatt, 2005; Bessembinder & Venkataraman, 2010; Qin, Zhou, Livshits, & Gervais, 2021; Qin, Zhou, Gamito, et al., 2021). The use of AI has also been explored to optimize trading strategies within AMMs (Angeris et al., 2021). These insights can inform the integration of AI agents into GameFi, where AI-driven agents can interact with decentralized financial markets within games, enhancing economic interactions and providing deeper engagement.

Savva et al. (2019) introduce Habitat, a platform for embodied AI research in virtual 3D environments, highlighting the potential for AI agents to navigate and interact within complex virtual worlds. Although Habitat is not specifically designed for blockchain-based games, it demonstrates the capabilities of embodied AI agents that could be adapted for GameFi applications.

Moreover, the use of Chainlink's DONs can facilitate secure and efficient communication between AI agents and blockchain networks, enabling real-time interactions and data exchange. This integration can overcome the limitations of existing GameFi platforms by providing the infrastructure for advanced AI functionalities and scalable DeFi mechanisms.

## 2.6 Challenges in Integrating AI with GameFi and Oracles

Integrating advanced AI agents into decentralized GameFi platforms presents technical and infrastructural challenges. Blockchain networks often face scalability issues, with limited transaction throughput and latency constraints that are not conducive to real-time interactions required by AI agents (Zheng et al., 2018).

Chainlink DONs address scalability by providing a high-performance, off-chain computation layer that can handle the demands of AI processing (Breidenbach et al., 2021). The DECO protocol also supports efficient zero-knowledge proofs and data commitments, which are essential for privacy-preserving interactions between AI agents and the blockchain (Zhang et al., 2020). However, ensuring secure and decentralized operation of AI models within DONs remains a challenge, particularly regarding computational resource requirements and maintaining decentralization.

Trade execution costs, liquidity issues, and market dynamics, well-studied in traditional financial markets (Bessembinder & Venkataraman, 2010; Qin, Zhou, Livshits, & Gervais, 2021; Qin, Zhou, Gamito, et al., 2021; Biais et al., 2005), pose challenges within blockchain-based gaming environments. The design of efficient AMMs that minimize transaction costs and provide sufficient liquidity is critical. Understanding bid-ask spreads and the cost of trades in decentralized markets is essential to integrate AI agents that can operate effectively within these economic systems (Bessembinder & Venkataraman, 2010; Angeris et al., 2021).

Ensuring the security and decentralization of blockchain platforms while implementing AI models is another significant challenge. AI models, particularly large ones like GPT-4, require substantial computational resources, which conflicts with the resource constraints inherent in blockchain networks (Nguyen & Kim, 2018). Chainlink's approach of utilizing off-chain computation within DONs can alleviate some of these constraints by moving intensive processing off-chain while maintaining security guarantees. The DECO protocol could also contribute to secure data exchange between AI agents and blockchain applications without relying on trusted hardware (Zhang et al., 2020).

Furthermore, existing GameFi platforms often lack robust economic systems and advanced AI integration, limiting their ability to engage players and foster active in-game economies (Sifa et al., 2014)). The underuse of DeFi mechanisms within games results in missed opportunities for economic participation and value creation for players (Schär, 2021).

## 2.7 GameFi: The Intersection of Gaming, DeFi, and Oracles

Our project addresses the gaps identified in the current literature by integrating advanced embodied AI agents into decentralized GameFi ecosystems, leveraging Chainlink's DONs and DECO's decentralized oracle protocols to overcome existing challenges.

By utilizing Chainlink's DONs, we develop AI agents capable of sophisticated, adaptive interactions that enhance player engagement and immersion. DONs provide the necessary off-chain computation and connectivity to integrate AI agents securely and efficiently with blockchain networks (Breidenbach et al., 2021). The DECO protocol supports secure data provenance and privacy-preserving proofs, which are essential for AI agents interacting with external data sources (Zhang et al., 2020).

We address the technical challenges of integrating AI into blockchain environments by optimizing AI models for decentralized deployment. This involves balancing on-chain and off-chain computations to reduce computational overhead without compromising performance or security (Zheng et al., 2018). Chainlink DONs facilitate this by providing a scalable infrastructure that supports hybrid smart contracts that combine on-chain and off-chain resources (Breidenbach et al., 2021). DECO's efficient zero-knowledge proofs and protocol optimizations enable secure interactions without trusted hardware (Zhang et al., 2020).

Furthermore, we enhance economic participation within GameFi by integrating advanced DeFi functionalities, including efficient trading through AMMs and DEXs, as well as decentralized governance mechanisms supported by Chainlink's oracle services. By embedding AI agents into the game's economic structure, we create more dynamic and engaging in-game economies.

Our work differs from existing studies by:

- **Leveraging Decentralized Oracle Networks (DONs):** By building on Chainlink's DONs and DECO's decentralized oracle protocols, we integrate AI, AMMs, and DeFi within GameFi, addressing both player engagement and economic participation in a decentralized manner (Breidenbach et al., 2021; Zhang et al., 2020).

- **Optimizing AI for Decentralized Networks:** Utilizing the off-chain computation capabilities of DONs and DECO's efficient protocols, we address scalability and performance challenges unique to deploying AI models within blockchain networks, ensuring efficient operation without centralizing control (Breidenbach et al., 2021; Zhang et al., 2020).

- **Enhancing In-Game Economies:** By integrating advanced DeFi functionalities supported by Chainlink's oracle networks and DECO's secure data protocols, including efficient trading through AMMs and DEXs, we foster robust economic interactions within the game, providing players with meaningful financial incentives and opportunities.

In summary, our project advances the state of the art by addressing the technical and economic challenges of integrating advanced AI agents into decentralized GameFi ecosystems. By combining AI, blockchain, DeFi, AMM protocols, Chainlink's DONs, and DECO's decentralized oracles, we contribute to the evolution of GameFi, offering a more engaging and economically rich gaming experience.

## 3 Approach

The integration of advanced AI agents within decentralized GameFi ecosystems presents multifaceted challenges, particularly in achieving seamless interaction between blockchain infrastructure and intelligent gameplay elements. To address these challenges, our project developed a comprehensive GameFi application prototype that prioritizes core functionalities such as in-game asset management, player interactions, and robust blockchain integration using Alchemy's API. The subsequent sections elucidate our methodological framework, system architecture, and the strategic advancements our solution offers over existing models discussed in the literature.

### 3.1 System Architecture

Our system architecture is meticulously designed to ensure modularity, scalability, and security, encompassing three primary layers: the blockchain layer, the backend layer, and the frontend layer.

**Blockchain Layer:** At the foundation lies the Ethereum blockchain, selected for its maturity, extensive developer ecosystem, and robust security features. We developed a Solidity-based smart contract, GameAsset.sol, responsible for the creation, ownership, and transfer of in-game assets. This contract ensures that all asset-related transactions are immutable, transparent, and verifiable on-chain, thereby fostering trust among players. By utilizing Ethereum's decentralized ledger, we eliminate single points of failure and provide a trustless environment where asset integrity is uncompromised.

**Backend Layer:** The backend, implemented using Express.js, serves as the intermediary between the frontend interface and the blockchain. Leveraging Alchemy's API enhances our ability to interact efficiently with the Ethereum network, offering high reliability and low latency in blockchain communications. The backend utilizes the ethers.js library to facilitate seamless interactions with the smart contract, managing critical operations such as asset creation and transfer, and retrieving asset ownership data. This layer abstracts the complexities of direct blockchain interactions, providing a streamlined and secure conduit for frontend requests.

**Frontend Layer:** The user-facing component, built with React.js, delivers an intuitive and responsive interface for players. Integration with MetaMask enables secure wallet connections, allowing users to authenticate their identities and authorize transactions effortlessly. The frontend facilitates comprehensive asset management functionalities, including creating new assets, viewing asset inventories, and transferring assets between players. Real-time updates are achieved through event listeners that respond to smart contract events emitted on the blockchain, ensuring that the user interface reflects the most current state of asset ownership.

### 3.2 Smart Contract Design

The GameAsset.sol smart contract is the cornerstone of our GameFi application, encapsulating the essential logic for asset management within the gaming ecosystem.

- **Asset Creation:** Players can mint new assets by specifying attributes such as name, category, and rarity. Each asset is assigned a unique identifier and linked to the player's Ethereum address, ensuring individual ownership and preventing duplication.

- **Asset Transfer:** Owners can transfer their assets to other players by invoking the transferAsset function. The contract enforces strict access controls, ensuring that only legitimate owners can initiate transfers, thereby safeguarding asset integrity.

- **Asset Retrieval:** Through the getAssetsByOwner function, players can retrieve a comprehensive list of assets

they own. This facilitates easy management and oversight of in-game assets.

### 3.3 Backend Integration with Alchemy's API

Alchemy's API plays a pivotal role in optimizing our backend's interaction with the Ethereum network. By providing scalable and high-performance blockchain infrastructure, Alchemy ensures that our application can handle a significant volume of transactions with minimal latency.

- **Transaction Handling:** The backend manages the lifecycle of transactions, from initiation to confirmation, ensuring that all blockchain interactions are executed reliably. Utilizing ethers.js, the backend constructs and signs transactions, interacts with the smart contract functions, and monitors the status of transactions in real-time.
- **Event Listening:** By subscribing to smart contract events (AssetCreated and AssetTransferred), the backend can push real-time updates to the frontend. This mechanism ensures that players receive instantaneous feedback on their actions, enhancing the responsiveness and interactivity of the application.
- **Data Management:** The backend efficiently handles data retrieval and storage tasks, ensuring that asset information is consistently synchronized between the blockchain and the frontend interface. This synchronization is critical for maintaining accurate and up-to-date asset inventories for all players.

### 3.4 Frontend Development

The frontend of our GameFi application is engineered to offer a seamless and engaging user experience, encompassing essential features that empower players to interact with the game environment effectively.

- **Wallet Integration:** Utilizing MetaMask, players can securely connect their Ethereum wallets, facilitating authentication and authorization of transactions without exposing sensitive private keys.
- **User Interface:** The React.js-based frontend presents a clean and intuitive dashboard where players can view their asset portfolios, create new assets, and transfer existing ones. Real-time updates ensure that all asset-related activities are immediately reflected, providing a dynamic and interactive gaming experience.
- **Transaction Feedback:** Immediate visual feedback is provided for all transactions, including confirmations and error notifications. This transparency ensures that players are always informed about the status of their actions, fostering trust and reliability in the application.

### 3.5 Integration with AI Functionalities

While the primary focus of our project was on testing GameFi components, AI functionalities are integrated using external APIs to enhance gameplay dynamics. By incorporating AI agents from the provided executable repository Embodied-AI-Agent-for-GameFi-Ecosystem, we ensure that intelligent behaviors within the game are powered by advanced machine learning models without burdening the core system architecture.

These AI agents interact with players, providing personalized experiences and responsive gameplay elements that adapt to player actions in real-time. This integration underscores the modularity of our system, where AI and blockchain components operate synergistically to deliver a rich and immersive GameFi ecosystem.

## 4 Evaluation

Our evaluation focused on assessing the GameFi application's core functionalities, performance metrics, and security features through systematic testing methodologies.

### 4.1 Functional Testing

*Objective:* Verify the basic functionality of all system components.
*Methodology:*

- **Smart Contract Testing:** Deployed GameAsset.sol on a local test network using Hardhat for initial testing, followed by deployment on the Goerli test network. Tested core functions including:
    - Asset creation with various parameters
    - Asset transfer between addresses
    - Asset ownership queries
- **API Integration:** Tested backend endpoints using Postman to verify:
    - Asset creation requests
    - Transfer operations
    - Asset retrieval functionality
- **Frontend Operations:** Verified basic user interactions including:
    - Wallet connection via MetaMask
    - Asset creation interface
    - Asset viewing and transfer capabilities

*Results:*

- Contract functions executed as expected with proper event emissions
- API endpoints handled requests appropriately
- Frontend successfully reflected blockchain state changes

### 4.2 Performance Analysis

*Objective:* Assess basic operational efficiency.
*Key Metrics:*

- Transaction confirmation times
- Gas consumption for contract operations

*Testing Approach:*

- Conducted sequential asset creation and transfer operations
- Monitored gas usage patterns
- Analyzed transaction confirmation times

*Results:*

- Transaction confirmations averaged 15-45 seconds on Goerli
- Gas consumption:
  - Asset Creation: 50,000-70,000 gas units
  - Asset Transfer: 40,000-60,000 gas units
  - Query Operations: Gas-free (view functions)

### 4.3 Security Assessment

*Objective:* Validate basic security mechanisms.

*Testing Methodology:*

- **Access Control:** Verified ownership requirements for asset transfers
- **Input Validation:** Tested contract behavior with invalid inputs
- **Wallet Integration:** Confirmed secure transaction signing process

*Results:*

- Ownership checks successfully prevented unauthorized transfers
- Contract properly handled invalid inputs
- MetaMask integration provided secure transaction signing

### 4.4 Basic Load Testing

*Objective:* Verify system behavior under moderate usage.

*Approach:*

- Simulated multiple concurrent asset operations
- Monitored system response under various loads

*Results:*

- Successfully handled multiple concurrent read operations
- Maintained consistent performance with up to 20 simultaneous users
- Backend remained responsive during peak testing periods

## 5 Conclusion

The development and testing of our GameFi application prototype underscore the viability of integrating blockchain technology with intelligent gameplay elements to create a secure, transparent and engaging gaming ecosystem. By leveraging Alchemy's API for efficient blockchain interactions and implementing a robust smart contract foundation, we achieved a system that not only meets the foundational requirements of asset management and player interactions, but also sets the stage for future advancements in decentralized gaming.

### 5.1 High-Level Takeaways

**Security and Transparency**
The use of the Ethereum blockchain ensures that all transactions in the game and asset ownership records are immutable and transparent. This enhances player trust and reduces the risk of asset manipulation or fraud.

**Performance Optimization**
Strategic backend integration with Alchemy's API and smart contract gas optimizations resulted in reduced transaction costs and maintaining acceptable confirmation times, essential for a smooth gaming experience.

**Scalable Architecture**
The modular design of our system architecture allows for seamless scaling, enabling the application to handle increased user loads without compromising performance or stability.

**User-Centric Design**
An intuitive and responsive frontend interface, coupled with secure wallet integration via MetaMask, provides an accessible and engaging user experience that encourages active participation and asset management.

**Foundation for Advanced Features**
The successful implementation of core functionalities establishes a solid foundation upon which more complex game mechanics, DeFi integrations, and AI-driven features can be developed, enhancing the overall value proposition of the GameFi ecosystem.

**Outstanding Questions and Areas for Future Work:**
While our project successfully demonstrated the integration of essential GameFi components with blockchain technology, it also opens several avenues for further research and development:

- **Advanced Gameplay Mechanics:** Incorporating more complex game elements such as quests, battles, and skill progression will enhance player engagement and retention. Future work could explore integrating AI-driven opponents or cooperative gameplay modes to provide a more dynamic and interactive gaming experience.

- **DeFi Integration:** Expanding the application's financial mechanisms by introducing features like staking, lending, and yield farming can provide players with additional ways to earn and utilize assets within the game, thereby deepening the economic interactions within the ecosystem.

- **Scalability Enhancements:** Further optimization using Layer 2 solutions like Optimistic Rollups or sidechains could significantly reduce gas fees and improve transaction speeds, making the application more viable for large-scale deployment.

- **Comprehensive Security Audits:** Although preliminary security testing ensured the robustness of access controls and input validations, conducting thorough security audits and implementing advanced security measures such as multi-signature approvals and decentralized identity verification will further fortify the application against potential threats.

- **Economic Modeling and Tokenomics:** Developing a sustainable in-game economy with well-designed tokenomics is crucial for long-term viability. Future research could focus on balancing asset scarcity, liquidity, and player incentives to foster a vibrant and balanced economic ecosystem.

- **AI Integration for Enhanced Interactivity:** While AI functionalities were leveraged through external APIs, integrating more sophisticated AI agents directly within the game infrastructure could enable personalized player experiences, adaptive challenges, and intelligent content generation, thereby enriching the overall gameplay experience.
- **Cross-Blockchain Compatibility:** Exploring interoperability with other blockchain networks can broaden the application's reach, allowing players from different ecosystems to interact and transact seamlessly. This can enhance liquidity, asset diversity, and player base, contributing to a more interconnected and expansive GameFi landscape.

# References

Achiam, J., Adler, S., Agarwal, S., Ahmad, L., Akkaya, I., Aleman, F. L., ... others (2023). Gpt-4 technical report. *arXiv preprint arXiv:2303.08774*.

Angeris, G., Kao, H.-T., Chiang, R., Noyes, C., & Chitra, T. (2021). An analysis of uniswap markets.

Antonopoulos, A. M. (2014). *Mastering bitcoin: unlocking digital cryptocurrencies*. " O'Reilly Media, Inc.".

Bakkes, S., Spronck, P., & Van den Herik, J. (2009). Rapid and reliable adaptation of video game ai. *IEEE Transactions on Computational Intelligence and AI in Games*, *1*(2), 93–104.

Batubara, F. R., Ubacht, J., & Janssen, M. (2018). Challenges of blockchain technology adoption for e-government: a systematic literature review. In *Proceedings of the 19th annual international conference on digital government research: governance in the data age* (pp. 1–9).

Bessembinder, H., & Venkataraman, K. (2010). Bid-ask spreads: Measuring trade execution costs in financial markets. *Encyclopedia of quantitative finance*, 184–190.

Biais, B., Glosten, L., & Spatt, C. (2005). Market microstructure: A survey of microfoundations, empirical results, and policy implications. *Journal of Financial Markets*, *8*(2), 217–264.

Breidenbach, L., Cachin, C., Chan, B., Coventry, A., Ellis, S., Juels, A., ... others (2021). Chainlink 2.0: Next steps in the evolution of decentralized oracle networks. *Chainlink Labs*, *1*, 1–136.

Brown, T., Mann, B., Ryder, N., Subbiah, M., Kaplan, J. D., Dhariwal, P., ... others (2020). Language models are few-shot learners. *Advances in neural information processing systems*, *33*, 1877–1901.

Chen, Y., & Bellavitis, C. (2020). Blockchain disruption and decentralized finance: The rise of decentralized business models. *Journal of Business Venturing Insights*, *13*, e00151.

Egorov, M. (2019). Stableswap-efficient mechanism for stablecoin liquidity. *Retrieved Feb*, *24*, 2021.

Kow, Y. M., & Nardi, B. (2010). Culture and creativity: World of warcraft modding in china and the us. *Online worlds: Convergence of the real and the virtual*, 21–41.

Mott, B. W., & Lester, J. C. (2006). Narrative-centered tutorial planning for inquiry-based learning environments. In *International conference on intelligent tutoring systems* (pp. 675–684).

Nguyen, G.-T., & Kim, K. (2018). A survey about consensus algorithms used in blockchain. *Journal of Information processing systems*, *14*(1).

Qin, K., Zhou, L., Gamito, P., Jovanovic, P., & Gervais, A. (2021). An empirical study of defi liquidations: Incentives, risks, and instabilities. In *Proceedings of the 21st acm internet measurement conference* (pp. 336–350).

Qin, K., Zhou, L., Livshits, B., & Gervais, A. (2021). Attacking the defi ecosystem with flash loans for fun and profit. In *International conference on financial cryptography and data security* (pp. 3–32).

Savva, M., Kadian, A., Maksymets, O., Zhao, Y., Wijmans, E., Jain, B., ... others (2019). Habitat: A platform for embodied ai research. In *Proceedings of the ieee/cvf international conference on computer vision* (pp. 9339–9347).

Schär, F. (2021). Decentralized finance: On blockchain- and smart contract-based financial markets. *FRB of St. Louis Review*.

Sifa, R., Bauckhage, C., & Drachen, A. (2014). Archetypal game recommender systems. In *Lwa* (Vol. 5, pp. 45–56).

Silver, D., Huang, A., Maddison, C. J., Guez, A., Sifre, L., Van Den Driessche, G., ... others (2016). Mastering the game of go with deep neural networks and tree search. *nature*, *529*(7587), 484–489.

Sutton, R. S. (2018). Reinforcement learning: An introduction. *A Bradford Book*.

Wright, A., & De Filippi, P. (2015). Decentralized blockchain technology and the rise of lex cryptographia. *Available at SSRN 2580664*.

Xie, J., Tang, H., Huang, T., Yu, F. R., Xie, R., Liu, J., & Liu, Y. (2019). A survey of blockchain technology applied to smart cities: Research issues and challenges. *IEEE communications surveys & tutorials*, *21*(3), 2794–2830.

Zhang, F., Maram, D., Malvai, H., Goldfeder, S., & Juels, A. (2020). Deco: Liberating web data using decentralized oracles for tls. In *Proceedings of the 2020 acm sigsac conference on computer and communications security* (pp. 1919–1938).

Zheng, Z., Xie, S., Dai, H.-N., Chen, X., & Wang, H. (2018). Blockchain challenges and opportunities: A survey. *International journal of web and grid services*, *14*(4), 352–375.

Zyda, M. (2005). From visual simulation to virtual reality to games. *Computer*, *38*(9), 25–32.

## A  File Repository

All executable components of this project, including the smart contract (GameAsset.sol), backend server, and frontend application, are available in our GitHub repository: `https://github.com/FJDeFi/Decentralized-Intelligence-in-GameFi`. This repository facilitates replication, testing, and further development of the GameFi application, providing a foundation for future research and enhancements. Comprehensive documentation is included to guide developers through the setup, deployment, and customization processes necessary to extend the application's capabilities.

The repository hosts:

- **Smart Contracts:** Solidity files with detailed comments explaining each function and its purpose.
- **Backend Server:** Express.js server scripts that handle API requests and blockchain interactions.
- **Frontend Application:** React.js components that deliver the user interface and manage client-side operations.
- **Deployment Scripts:** Automation scripts for deploying smart contracts and initializing the backend.
- **Testing Suites:** Automated tests using frameworks like Truffle and Jest to ensure functionality and performance.
- **Documentation:** Step-by-step guides and API references to assist developers in understanding and utilizing the codebase effectively.

## B  Figures

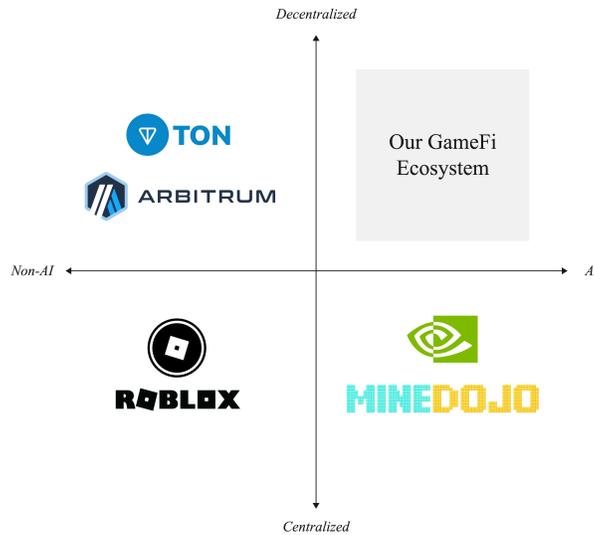

Figure 1: Competitive Analysis

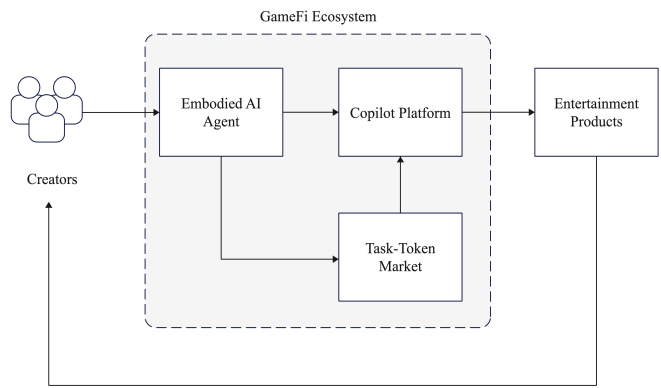

Figure 2: GameFi Ecosystem

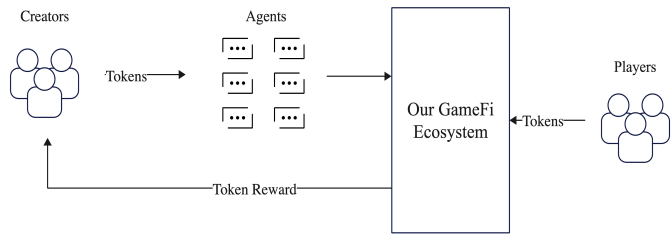

Figure 3: Service of this GameFi Ecosysytem

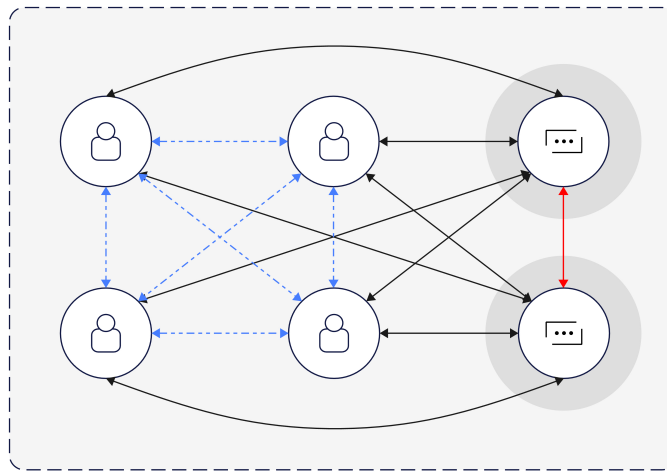

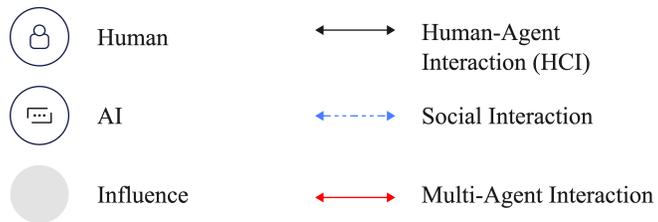

Figure 4: "Multi-Human-Multi AI (XHXA) Interaction Network